\documentclass[final,5p,times,twocolumn]{elsarticle}
\usepackage{graphicx,subfigure}
\usepackage{amsmath,amssymb}
\usepackage{bm}

\makeatletter
\def\ps@pprintTitle{%
 \let\@oddhead\@empty
 \let\@evenhead\@empty
 \def\@oddfoot{}%
 \let\@evenfoot\@oddfoot}
\makeatother

\newcommand{\be}{\begin{eqnarray}}
\newcommand{\ee}{\end{eqnarray}}

\def\slashchar#1{\setbox0=\hbox{$#1$}           
   \dimen0=\wd0                                 
   \setbox1=\hbox{/} \dimen1=\wd1               
  \ifdim\dimen0>\dimen1                        
 \rlap{\hbox to \dimen0{\hfil/\hfil}}      
  #1                                        
 \else                                        
    \rlap{\hbox to \dimen1{\hfil$#1$\hfil}}   
    /                                         
 \fi}                                         %

\journal{Nuclear Physics A}

\begin{document}

\begin{frontmatter}

\title{Classical  interactions of the instanton-dyons with antidyons\tnoteref{label1}}
 \tnotetext[label1]{}
 \author{ Rasmus Larsen \corref{cor1}}
 \cortext[cor1]{Corresponding author}
\ead{rasmus.n.larsen@stonybrook.edu }
\author{ Edward  Shuryak \corref{cor2}}

\address{Department of Physics and Astronomy, Stony Brook University,
Stony Brook NY 11794-3800, USA }

\title{Classical  interactions of the instanton-dyons with antidyons}

\author{}

\address{}

\begin{abstract}
Instanton-dyons, also known as instanton-monopoles or instanton-quarks, are topological constituents of the instantons at nonzero temperature and nonzero expectation value of $A_4$. 
While the interaction between instanton-dyons has been calculated to one-loop order by a number of authors,  
that for dyon-antidyon pairs remains unknown  even at the classical level. In this work we are filling this gap, by solving the gradient flow equation 
 on a 3d lattice. We start with two well separated  objects.  We find that, after initial rapid relaxation, the configurations follow ``streamline"
set of configurations, which is basically independent on the initial configurations used. In striking difference to
instanton-antiinstanton streamlines, in this case it ends at a quasi-stationary
configuration, with an abrupt drop to perturbative fields. We parameterize
the action of the streamline configurations, which is to be used in future many-body calculations.
\end{abstract}


\end{frontmatter}

\section{Introduction}

Instantons  \cite{Belavin:1975fg}  are Euclidean 4-dimensional topological solitons of the Yang-Mills gauge fields known to be important ingredients of the
gauge fields in the QCD vacuum, as well as at finite-temperatures comparable to the critical one $T\sim T_c$.  
As requested by a index theorem, the topological charge of the instantons leads to fermionic zero modes, which provide the so called 't Hooft interaction,
 explicitly violating the $U_A(1)$ chiral symmetry. At sufficient density of instantons, these zero modes
 are collectivized and
 create the so called Zero Mode Zone  of quasi-zero eigenstates, which breaks spontaneously the $SU(N_f)$ chiral symmetry. To describe those phenomena, 
 the so called Interacting Instanton Liquid Model (IILM) has been developed, which includes the
 instanton-induced  't Hooft interaction
 to all orders,  for a review  see
\cite{Schafer:1996wv}.   Although those states are only a tiny  subset of all fermionic states in lattice numerical simulations,  
 by removing them from the propagators one observes restoration of $SU(N_f)$ and $U(1)_A$ chiral symmetries and 
 a significant modification of hadronic masses, for recent works see e.g. \cite{Denissenya:2014ywa,Denissenya:2014poa}.

The first step toward   instantons at finite temperatures was finding the so called ``caloron" solution
\cite{Harrington:1978ve}, which is periodic in  Euclidean time.  
The second  -- and much more nontrivial -- step \cite{Lee:1998bb,Kraan:1998sn} included  the 
 nonzero mean value of the
4-th component of the gauge field $\langle A_4^3\rangle=v$. This so called ``KvBLL caloron" solution revealed the substructure of the (anti)instanton: at nonzero $v$ it gets split into $N_c$
(number of colors) separate  (anti)self-dual 3d solitons with nonzero (Euclidean) electric and magnetic charges. 
Those are in literature called the instanton-dyons, instanton-monopoles and instanton quarks: we will call
them below ``dyons" for short. While they are 't Hooft-Polyakov monopoles, with the role of adjoint scalar played
by $A_0$ component of the gauge field, the terminology changes slightly since the gradient of $A_0$
is naturally called the electric field. 

In this work we will focus on the simplest gauge group $SU(2)$:  it  has only one diagonal generator $\tau^3/2$ and thus a single
Abelian subgroup  remains unbroken by the nonzero VEV $<A_0^3>=v$. In this case  
there are  four instanton-dyons, corresponding to all possible combinations of electric and magnetic charges.
By tradition the time-independent selfdual ones are called $M$ with charges $(e,m)=(+,+)$ and 
time-twisted $L$ with charges $(e,m)=(-,-)$, the anti-selfdual antidyons are called  
 $\bar{M}$, $(e,m)=(+,-)$ and  $\bar{L}$, $(e,m)=(-,+)$. 
 Other names used in literature ~\cite{Poppitz:2012nz}  are the ``BPS monopoles" for M and ``KK" ones for L.
 Recent example of a lattice work identifying those instanton-dyons
 is \cite{Bornyakov:2014esa}, for earlier work see references therein.

Recent studies of the  instanton-dyons had developed along two different but related
directions. One of them~\cite{Poppitz:2012nz} starts in  a very specific supersymmetric setting,
  with compactification to $R^3\otimes S^1$.  Small radius of a circle
 leads to a weak coupling regime, in which all  topological objects are  exponentially suppressed, yet under a theoretical control. Preservation of  supersymmetry requires $periodic$ fermions: if it is there, the perturbative Gross-Pisarski-Yaffe
potential is canceled, and even exponentially small density of the dyons and their pairs can lead to 
the confining value of the holonomy $v$. 
 
Another group\cite{Shuryak:2012aa,Faccioli:2013ja} study instanton-dyons in the pure gauge and/or QCD-like theories.
in which dyons are dilute at high temperature but form a rather dense plasma at $T\sim T_c$. It has been 
 argued \cite{Shuryak:2013tka} that sufficient density and repulsion between the instanton-dyons should result
 in  confinement, in spite of the  Gross-Pisarski-Yaffe
potential  working against it. 
Theories with quarks also have dyon-induced fermionic zero modes, and at sufficiently large dyon density the 
 breaking of the chiral symmetries takes place \cite{Faccioli:2013ja}.  So, the instanton-dyons are crucially important for 
understanding of the interrelation of confinement and chiral symmetry breaking, the two main nonperturbative
phenomena in QCD-like theories. 

In order to understand the dyon ensemble  quantitatively, one obviously needs to know first
 the
forces acting between them.
A principal difference between the (i) single duality sector
(only self- or antiselfdual objects) with (ii) the interaction between self- or antiselfdual objects
is that only in the former case does
  the  celebrated Bogomolny inequality becomes equality, requiring the action of the configuration to be
entirely determined by its global topological charge. This 
eliminates interaction at the classical $O(1/g^2)$ level in case (i),  producing the {\em moduli space} intensely studied in  mathematical literature.  The moduli space metric for those and related spaces
can be calculated, 
providing the measure of integration over the collective variables.
This metric is traditionally expressed via a determinant of a certain matrix.
For the LM pair this metric has been calculated by Diakonov et al. 
\cite{Diakonov:2004jn}: at large separation it reduces to one-loop  Coulomb-like interactions $O(1/r_{LM})$.
 Later Diakonov \cite{Diakonov:2009ln} conjectured  a volume element for any number of $M,L$ dyons,
combining the DGPS one with Gibbons-Manton approximation to   Atiyah-Hitchin  metric. This interaction has been used  in the  numerical simulations of the dyon ensemble \cite{Faccioli:2013ja} .

 The second case
(ii) -- the interaction of self and anti-selfdual objects -- is however much more difficult to study.
  There is no BPS protection of the action at the  classical level  and thus no moduli spaces or corresponding solutions.
 The aim of the present paper is to map the corresponding configurations via the so called  {\em  streamline}, a one-parameter set of configurations defined by a condition
 that the driving force $\delta S / \delta A_\mu$, while nonzero, is {\em tangent to the set}. 
   The practical way to generate them is to solve the {\em gradient flow equation}, starting from some initial ansatz,
   as e.g. was done long ago   for the instanton-antiinstanton  in the double-well potential  \cite{Shuryak:1987tr}. 

For gauge field instantons  this was done analytically by Balitsky and Yung \cite{Balitsky:1986qn} in the
large-distance limit.  Since classical 
 instanton-antiinstanton problem is intrinsically conformal, one can perform a conformal transformation into
a co-central configuration, relating the gauge theory  and the double-well instantons.
Using this method  Verbaarschot  \cite{Verbaarschot:1991sq}  found a quite accurate analytic
approximation to the instanton-antiinstanton streamline. Recently, after the influential paper by Luscher \cite{Luscher:2011bx},
the gradient flow method is now widely used, to determine the beta function and thermodynamical observables on the lattice.
In our paper, however, we remain in the classical setting, with non-running coupling, and so we will not
discuss these applications.

 In this paper we  study the streamline configurations of the  
 dyon pairs  with the same electric but opposite magnetic charges,
 namely $M\bar{M},  L\bar{L}$.
(Unfortunately, the finite $T$ problem at nonzero  $v$ is not
conformal, and there is no transformation into a co-central case available.) 
 Their total action  defines what one can call ``classical interaction potential".
 Although this interaction is parametrically larger than that following from (one loop) moduli spaces 
 or fermionic zero modes, it has been unknown and thus not included in the first simulations \cite{Faccioli:2013ja}:
 we certainly plan to do so in subsequent publications. 


\section{The setting}\label{Setting}
\subsection{Instanton-dyons and their superposition}
We do not present here extensive introduction on the configurations and their history, which can be found
e.g. in  \cite{Diakonov:2009ln}. 
%
Let us just remind that ``Higgsing" the SU(2) gauge theory by a nonzero VEV of $A_4$  splits three gluons into two massive
and one massless (diagonal)  one, according to which the  Abelian charges are defined.
In the simplest so called  hedgehog gauge, in which the color direction of the  ``Higgs" field  at large r is directed along the unit radial vector  
$A_4^m\rightarrow v \hat r_m$, the solutions are  
\begin{eqnarray}
A_4 ^a &=& \pm \hat{r}_a \left( \frac{1}{r} -v \coth (v r) \right) \nonumber \\
A_i ^a &=& \epsilon _{aij} \hat{r}_j \left(\frac{1}{r}-\frac{v}{\sinh  (v r)}\right), \label{Afield}
\end{eqnarray}
where $+$ corresponds to the $M$ dyon and $-$ corresponds to the $\bar{M}$ dyon. $r$ is the length in position space. The $L$ and $\bar{L}$ dyon are obtained by a replacement $v \to 2 \pi T -v$ and a certain time-dependent gauge change.


Any superpositions of the dyons at nonzero $A_4$ are   nontrivial 
since one should match at large distances not only in magnitude, but also its direction in color space.
Those can be achieved by the following four-step procedure:\\
(i) ``combing", or going to a gauge in which the ``Higgs field" $A_4^3=v$ (upper index is color generator, lower index is
the Lorentz one) at large distances is
the same in all directions and for all objects\\
(ii)  performing a time-dependent gauge transformation which removes  $v$ completely\\
(iii) superimposing the dyons in this gauge\\
(iv)  making reverse time-dependent gauge transformation, reintroducing $v$ .

(i) Description of the ``combing" procedure can be found in  \cite{Diakonov:2009ln}, but since 
 there are misprints in this reference we remind the main formulae.  The gauge
matrices are  rotations which put a radially directed unit vector into $\pm z$ direction. 
It is convenient to write those using spherical coordinates $r,\theta,\phi$ instead of Cartesian coordinates $x$.
The plus one  is
\[\arraycolsep=1.4pt\def\arraystretch{2.2} S_+(x) = \left( \begin{array}{cc}
cos( {\theta \over 2}) & sin({\theta \over 2}) e^{-i\phi} \\
-sin({\theta \over 2}) e^{i\phi} & cos( {\theta \over 2}) \end{array} \right),\]

 $S_-$ is obtained by setting $\theta \to \pi-\theta$. It should be noted that this choice of transformation is not unique. The matrix $\Omega=S_\pm$ is used in the general 
gauge transformation of the gauge field  
\be  {A_\mu=>  A}_\mu'=\Omega A_\mu \Omega^\dagger   -  i (\partial_\mu \Omega) \Omega^\dagger   ,   \ee
which is expressed in a standard matrix-valued  form 
\be A_\mu= A^a_\mu {\tau_a \over 2},\ee
where Pauli matrices divided by two are the  SU(2) generators in standard normalization.

(ii) The next gauge rotation matrix depends on Euclidean time and is 
\be \Omega_2= exp( -i x_4 v {\tau_3 \over 2}), \ee
so the derivative term produces $-v$ and cancels the original expectation value. 

(iii) The rotated dyon and antidyon are simply added together
\be A_\mu= A_\mu^{dyon}+A_\mu^{antidyon}.  \ee

(iv) Now one has to perform a gauge rotation, $opposite$
 to that in point (ii), with $\Omega_3=\Omega_2^+$. 
Since these rotations commute, they just cancel each other except for the derivative term
which puts back  $v$ at infinity. (If one would not perform steps (ii) and (iv)
but would naively do step (iii), the expectation value of $A_4^3$   would be $2v$.)

Superimposing two dyons by such a procedure, a sum of the correctly combed potentials, is what we call {\em 
the sum ansatz}. 
Needless to say, it is  an approximate solution only at large separation between the solitons,
used only as the starting point in our studies. Before we put these configurations on the lattice, as we detail shortly, 
we calculated the corresponding 
 fields strengths and currents in Cartesian coordinates, both in Maple and Mathematica\footnote{The reader should be warned that one has to redefine the inverse
trigonometric functions with the right branches from those in the default setting, to get correct fields.}.

As is well known, a ``combed" monopole or dyon  possesses the Dirac string, a singular
gauge artifact propagating one unit of magnetic flux from infinity to the dyon center. 
By selecting an appropriate gauge one can direct the Dirac string to  an arbitrary direction.
Superimposing two dyons with different directions of the Dirac string, one gets  {\em non-equivalent}
configurations: the interference of singular and regular terms make the Dirac strings no longer invisible or pure gauge
artifact. (However, this is cured during the gradient flow process, as we will discuss below.)

\begin{figure}[t]
  \begin{center}
      \includegraphics[width=7cm]{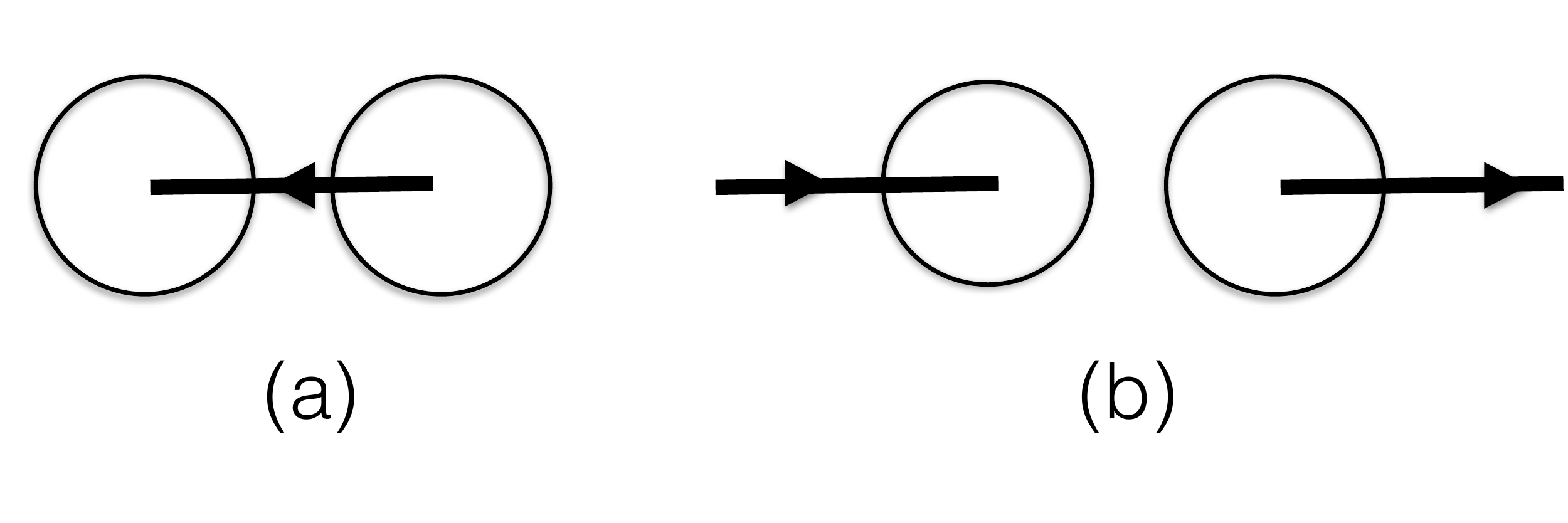}
   \caption{Two extreme positions for the Dirac strings, for  the $M\bar{M}$ pair.  
  \label{fig_Dstrings} }
  \end{center}
\end{figure}

 Two  extreme selections for the  Dirac strings are: (a) 
 a ``minimally connected dipole" when it goes along the  line
connecting two dyon centers;  and (b)  a ``maximally disconnected" pair,
in which two  Dirac strings approach two  centers from the opposite directions, see Fig.\ref{fig_Dstrings}.
Under the gradient flow the former is supposed to reach magnetically trivial configuration,
while the latter relaxes to a (pure gauge) Dirac-string-like state passing the flux through the system, from minus to plus infinity.
The former case appears to be simpler: but our experience has shown  
that the type-(b) configuration generates smaller artifacts, since the Dirac strings interfere less. We will take the type-(b) 
configuration as our starting configuration.

   The sum ansatz possesses certain artifacts, e.g.
   the Dirac strings become visible in the action plot. This is to be expected 
   due to interference of the singular Dirac string with regular solution for the other 
   dyon.  Furthermore, a correct smooth behavior at the center of each dyon is
also violated, as well as a left-right symmetry between the dyon and antidyon.
To cure some of the artifacts  one may invent certain improved profiles. For example
multiplying the ``Higgs" component of the field  by the factor 
\be A_4^3\rightarrow A_4^3  { (x-X_{M})^2 (x-X_{\bar{M}})^2  \over [\rho^2+ (x-X_{M})^2 ][\rho^2+ (x-X_{\bar{M}})^2 ]} 
\label{ans_mod},\ee 
 which  forces the field to  vanish at the  centers.
However, we observed that the gradient flow procedure eliminates such artifacts automatically,
with results quite independent of the shape of the starting configuration, so no such improvements 
are actually needed.

 \subsection{The gradient flow}
 
 The ``force" driving gradient flow is the current
 \be
	j_\mu^a \equiv -\frac{\delta S}{\delta A_\mu^a}|_{A=A_\textrm{ansatz}}=(D_v^{ab}G_{v\mu}^b)|_{A=A_\textrm{ansatz}}\neq0.
\ee
For solutions of the YM equation, such as a single dyon, it
vanishes at all points.  
For dyon-antidyon configurations which
we study it is nonzero, showing    the direction of the gradient flow 
towards the reduction of the action.   





Introducing the computer  time  $\tau$ we can write the trajectory of the resulting gradient flow  according to the equation 
\be\label{eq:steepdesc}
	\frac{\textrm{d}A_\mu^a}{\textrm{d}\tau}=-\frac{\delta S}{\delta A_\mu^a}.
\ee


\section{Dyons on the lattice}\label{lattice}
\subsection{The gauge fields}
On the lattice the representation of the gauge field is given in terms of the so-called link variables
\be
	U_\mu(x)\equiv \mathcal{P}e^{ig\int_x^{x+\hat{e}_{\mu}}A_\mu(z)\textrm{d}z}=e^{igaA_\mu(x+\hat{e}_{\mu}/2)}+\mathcal{O}(a^3)
\ee
where $a$ is a lattice spacing, assumed small, and 
\be
	U_{-\mu}(x)=U^\dag_\mu(x-\hat{e}_{\mu}).
\ee

The simplest gauge invariant quantity we can build using the gauge link is the plaquette
\be
	P_{\mu\nu}(x)=U_\mu(x)U_\nu(x+\hat{e}_{\mu})U^\dag_\mu(x+\hat{e}_{\nu})U^\dag_\nu(x)
\ee
and with the plaquette we can define a lattice gauge action with the correct continuum limit: $S=\frac{1}{4}\int\textrm{d}^4x\, F^{\mu\nu\, a}F_{\mu\nu}^a$
\be
	S=\frac{2 N}{g^2}\sum_x\sum_{\mu<\nu}\left(1-\frac{1}{2N}\textrm{Tr}[P_{\mu\nu}(x)+P^\dag_{\mu\nu}(x)]\right).
\ee

To visualize the gauge field it will be useful to plot the action density using
\be
	s(x)=\frac{2 N}{g^2}\left(1-\frac{1}{48N}\textrm{Tr}\left[\sum_{\substack{\mu,\nu=\pm1\\ \mu<\nu}}^{\pm4}(P_{\mu\nu}(x)+P^\dag_{\mu\nu}(x))\right]\right).	
\ee

Let us now  translate eq. (\ref{eq:steepdesc}) into the lattice language.

All the transformations explained in section \ref{Setting} will thus be performed on the lattice, with the link gauge transformations 
\begin{eqnarray}
U_\mu(x) &\to & \Omega(x) U_\mu(x) \Omega ^\dagger(x+\hat{e}_\mu). 
\end{eqnarray}
We still call it a  ``sum ansatz" although the terms are now multiplied instead. (i) We first comb the matrix $U_4$ by rotating it, such that $U_4$ has no $\tau _1$ or $\tau _2$ component. (ii) We do a gauge transformation in time to make the asymptotic value of $U_4$ equal to the identity matrix $I$. (iii) We multiply the two gauged dyon configurations. (iv) We do another gauge transformation to reintroduce the right value of $\tau _3$ for the asymptotic value of $U_4$. This leaves an extra term when we add the two dyons given as the following element of the temporal gauge transformation
\begin{eqnarray}
\delta \Omega_t &=& \exp(iav\frac{\tau _3}{2}).
\end{eqnarray}
The time dependent parts cancel when we reintroduce the asymptotic value of $A_4$, leaving $\delta \Omega_t$ behind. $\delta \Omega_t$ is not present in $U_i$ since the gauge transformation that remove the asymptotic value of $A_4$ is simply $\Omega(x) U_i(x) \Omega ^\dagger(x)$. We therefore end with the $SU(2)$ matrices given by
\begin{eqnarray}
&&U_4(x) = \\
&& S_ +(x) U_{1,4}(x) S_ +^\dagger(x+\hat{e}_4)\delta \Omega_t S_ -(x) U_{2,4}(x) S_ -^\dagger(x+\hat{e}_4) \nonumber \\
&&U_i(x) = \\
&& S_ +(x) U_{1,i}(x) S_ +^\dagger(x+\hat{e}_i)S_ -(x) U_{2,i}(x) S_ -^\dagger(x+\hat{e}_i), \nonumber 
\end{eqnarray}
where $U_{1,\mu}(x)$ and $U_{2,\mu}(x)$ are the links of the $\bar{M}$ and $M$ dyon given in equation \ref{Afield}. It should be noted that the gauge transformation $S$ is defined around the dyon it combs. 

All time dependence is canceled, so we decide to work in 3 dimensions only, since the gradient flow will be the same for all times.

For the $M\bar{M}$ configuration we comb $M$ with $S_-$ and $\bar{M}$ with $S_+$. The initial configuration for $L \bar{L}$ is similar. Here we comb $L$ with $S_+$ and $\bar{L}$ with $S_-$. This means that the asymptotic value of $A_4$ becomes negative instead.


Varying the action with an infinitesimal SU(2) rotation $U_\mu (x) \to \left(I+i\tau _k \epsilon _k\right)U_\mu (x)$ one finds the standard  current expression 
\begin{eqnarray}
J_\mu (x) &=& \sum_{\nu}\left(P_{\mu\nu}(x)-P^\dag_{\mu\nu}(x)\right)\\
 & - &\sum_{\nu}\left(P_{\mu\nu}(x-\hat{e}_v)-P^\dag_{\mu\nu}(x-\hat{e}_v)\right), \nonumber
\end{eqnarray} 
where plaquettes $P$ should be understood as the product of 4 links, always started from the same point $x$ (as needed for correct gauge covariance)
 in the direction $
\mu$. All plaquettes that contain $U_\mu(x)$ come with a plus sign and all plaquettes that contain $U  ^\dagger_\mu(x)$ come with a minus sign. The next step is a projection onto the $SU(N_c)$ color generators  
\begin{eqnarray}
J_{i,\mu}\equiv dt\textrm{Tr}[i\tau _i J_\mu (x)]. 
\end{eqnarray}
which eliminates possible contribution proportional to the unit matrix.

The matrix used for actual updates of the link variables is calculated as
\begin{eqnarray}
L_\mu(x) &=& \sqrt{J_{1,\mu}^2+J_{2,\mu}^2+J_{3,\mu}^2}\\
\theta _{i,\mu}(x) &=& J_{i,\mu}/L_\mu\\
C_\mu(x) &=& \cos(L_\mu)I+i\sin(L_\mu)\sum _i \theta _{i,\mu}\tau _i .
\end{eqnarray}
The multiplication of all links by 
$C_\mu(x)$ 
\begin{eqnarray}
U_\mu (x) &\to & C_\mu(x)U_\mu (x).
\end{eqnarray}
is our version of one step of the gradient flow. We checked that, with the double precision code used,
the link matrices remain belonging to $SU(2)$ within small errors, even after thousands of time steps needed in
the calculation. 
 For small enough $d\tau$ the action should monotonously decrease,
and indeed it does, at all locations and at all times.

\subsection{ Lattice details}\label{lattice_details}
Since $M\bar{M}$ pairs and $L\bar{L}$ pairs are time independent ($L\bar{L}$ pairs are time independent in the gauge where the Higgs field is $-(2\pi T -v)$), the lattice used is three-dimensional with size $N^3$. The fields on it are not periodic.   In order to protect the expectation value of $A_4^3$ during gradient flow, we hold the sides of our cube constant, i.e. we don't update the links  on the edges of the lattice. 

Most of the calculations are done with  $64^3$ cubic lattice. Its 
 size in absolute units is $40/v$ in each dimension with $a=0.625/v$, unless otherwise specified. 
This might seem like a rough lattice. While all  configurations {\em before} combing 
 have sufficiently small $A$ even at  the cores,  so that $|aA_\mu| \ll 1$, 
 $after$ combing    large fields $a A\sim 1$ do appear, coming from the Dirac string:
 however those are pure gauge and 
 they do  not affect the action at the streamline part of the process, as we will explain below.

On this setup the discretized analytic solution of one dyon is stable under gradient flow. Its action is  $5\%$ lower than the analytic value of $4\pi v$, which is due to fields outside of our box. The
  absolute value of electric and magnetic charge, calculated by the Gauss  flux integrals over certain cubes near the box surface, are both equal to $\pm 1$, inside the numerical accuracy of the double precision code we use.

The 4-d gauge action  expressed in terms of the 3-dimensional action is \be S={1\over g^2}\int_0^{1/T} dx_4 S_3={S_3 \over g^2T } \ee
which is itself dimension full and  scales as  $S_3\sim v$: thus the $M$ dyon action is $S\sim v/g^2 T$.
The actual value of $T$ and the gauge coupling $g$ are irrelevant for our calculation of $S_3$ 
since it is just an overall
 factor in the action $S$. Furthermore,
since our classical 3d theory is scale invariant  $A_\mu \to vA_\mu$ and $r\to vr$, the absolute units of $v$ are
unimportant and  we can use $v =1$.  In other words,  all distances are  in units of $r*v$.

Apart from the action and electric and magnetic charges, we also monitor the presence of  the Dirac strings as the system evolves. 
The circulation integrals $\oint dx_\mu A_\mu$  around the Dirac string are calculated,
 by adding the phases of subsequent links in the $\tau _3$-direction, 
using the inverse of the parameterization 
\begin{eqnarray}
U_\mu (x) = \cos(\phi)I+i\sin(\phi)\sum _i \theta _{i,\mu}\tau _i.
\end{eqnarray}
We do observe the famous $2\pi$ phase circulation at all times of the gradient flow, indicating that the
Dirac string flux through our box remains there.
 
 \section{Results}
\subsection{Qualitative features of the streamline}

Before we present our results in detail, we would like to give a brief overview of the findings,
starting with
 a reminder of the streamline  for the instanton-antiinstanton case.
 These configurations, either in quantum mechanical setting    \cite{Shuryak:1987tr} or gauge fields \cite{Verbaarschot:1991sq},
 have the topological charge zero and a meaning of tunneling forth and back, with only finite time spent in the second well (valley). When this time is small, there
 is no reason for the configuration itself to be different from zero (path or gauge fields). So,  the end of the 
 instanton-antiinstanton  streamline are
 the configurations with a small action $S\sim 1$ which cannot be treated semiclassically.  
 
 The case under consideration, with the instanton-dyons, is quite different. While two charges -- the magnetic and the topological ones -- still add to zero and
 can annihilate each other,  there is one remaining -- electric -- charge, which adds to 2. 
 Our definition of charge is based on the flux through a closed surface, and the charge density is the divergence of the field. Since the electric charge is not conserved in this definition, there is no reason that the electric charge has to be equal to 2 throughout the streamline.
 

 One might think that
 the process is dominated by the electric charge. We found it is not the case, and it is the
 behavior of the magnetic charge which is most important.
 
 The gradient flow process was found to proceed via the following stages: \\
 (i) {\em near initiation}: starting from an ansatz described above one finds rapid reduction of the action and
 disappearance of artifacts related with the Dirac strings\\
 (ii) relatively slow and universal evolution along the {\em streamline set}. The action decrease is small but steady. The dyons basically approach each other, with relatively small deformations:
 thus the concept of an interaction potential between them makes sense at this stage\\
 (iii) a {\em metastable state} at the streamline's end: the action remains practically constant, evolution is very slow and 
 is an internal deformation of the dyons rather than their further approach\\
   (iv) {\em rapid collapse} into  perturbative fields, plus some zero action (pure gauge) remnants  \\   
 
\begin{figure}[h!]
\includegraphics[scale=0.48]{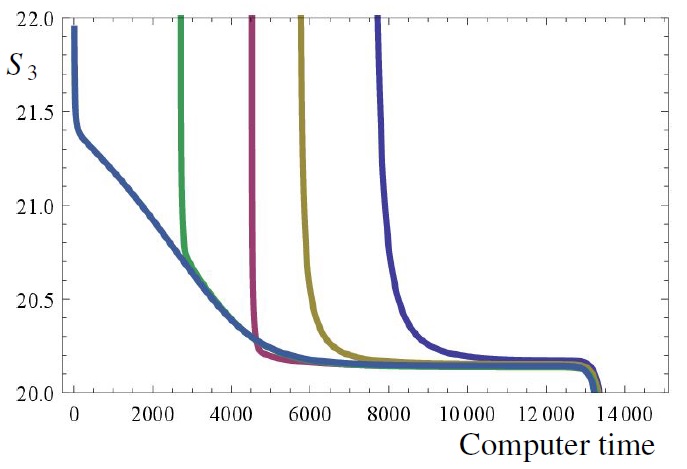}
\caption{3d-action $S_3$ for $v =1$ as a function of computer time for an initial separation $|r_M-r_{\bar{M}} |v=0$, $2.5$, $5$, $7.5$, $10$ between the $M$ and $\bar{M}$ dyon from right to left in the graph. The action of two well separated dyons is 23.88 for the lattice with $64^3$ points.}
\label{Streamlines}
\end{figure}

A sample of computer time histories for the total action is shown in  Fig.\ref{Streamlines}.
The stage (i) corresponds to near-vertical initial evolution, stage (ii) to declining universal line, stage (iii) to the
horizontal part at the right, following by another vertical line of total action collapse to zero (not shown).


Crucially important  is the observation that, even at the end of the streamline, the  action value is not that far from the sum of those of the two separated dyons. In other words, the
classical interaction potential we found is in a sense {\em numerically small}. 

We  observe an $universality$ of the streamline:  independent on the initial ansatz and even 
initial dyon separation we find that our gradient flow proceeds
through essentially the same set of configurations at stages (ii-iv).   A parameter 
we found most practical for their  characterization
is simply their $lifetime$ -- duration, in our computer time $\tau$, from a particular configuration to the  final collapse. 
To emphasize that, in Fig. \ref{Streamlines} we have drawn histories with different initial but the same final times.

The existence of stage (iii)  has not been anticipated. All configurations corresponding to it
have the same action, and -- within our accuracy the same dyon-antidyon distance. 
One can perhaps lump all of them into a new metastable configuration,  a dyon-antidyon molecule. (Perhaps
those can be  identifiable in the lattice gauge field ensembles.)

For configurations with the initial  separation smaller than $4.2/v$, 
we observe that dyons move away from each other, to the same metastable configuration.

\subsection{ Parameterization of the $M$ and $\bar{M}$ Streamline}

\begin{figure}[h!]
\includegraphics[scale=0.48]{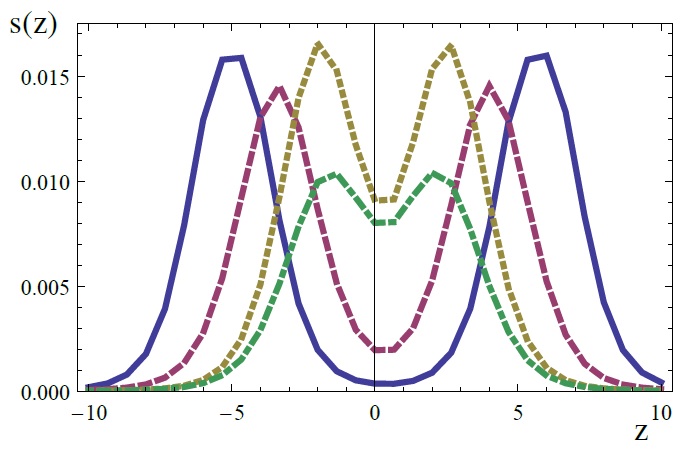}
\caption{Action density along the z axis in natural units for a separation $|r_M-r_{\bar{M}} |v=10$ between the centers of the two dyons. The configuration with the maximums furthest from each other is the initial configuration. At computer time $\tau= 3000$ it has moved further towards the center. At $\tau=12000$ the configuration has reached the metastable configuration with a separation between the maximums of $4.2/v$. At $\tau=13700$ the configuration has collapsed to a single maximum, which  continues to shrink until the action vanishes. Histories  shown
correspond to those displayed in Fig. \ref{Streamlines}. }
\label{Action_density}
\end{figure}

To define the ``interaction potential" between dyons and antidyons  we need two things. First, we need the action $S_3$ as a function of computer time, a sample of which was already shown in Fig. \ref{Streamlines}. 
 Second we need to define the separation between the dyons, and follow it as a function of computer time
 over the gradient flow. Locations of the 
 dyons at a specific computer time is inferred from the two maxima of the action density. We define it in each configuration by fitting 3 points around each maximum with a second order polynomial. 
 A sample of  action density distribution is shown in Fig. \ref{Action_density}.   
 
Combining the actions and dyon separations  we obtain 
the interaction potential. We use the configuration that starts with a separation of $10/v$ between the dyons to obtain the interaction potential, since it was the configuration that started with the largest separation. The range of 
 separations,  as always in  units of $1/v$, is from (slightly smaller than) $r=10/v$ to $r=4.2/v$:
 at this last value the configurations collapse to pure gauge with zero action.
 
\begin{figure}[h!]
\includegraphics[scale=0.49]{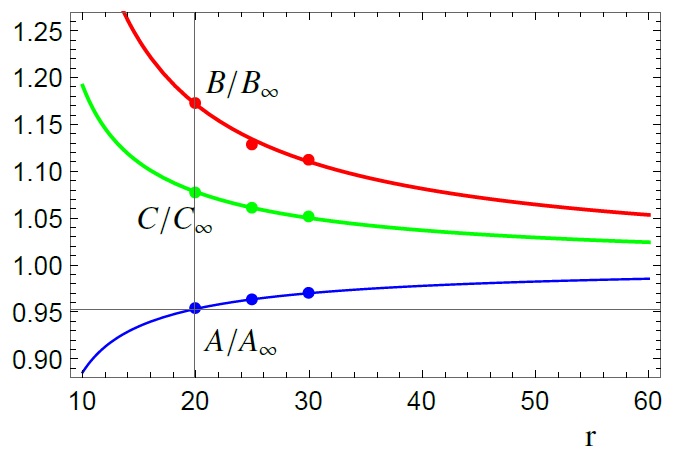}
\caption{The 3 parameters $A$, $B$ and $C$ normalized by their value at $r=\infty$ ($A_\infty$, $B_\infty$ and $C_\infty$) as a function of lattice half width r at $v a=40/64$. We extrapolate using $c+b\left[h(r)-h(\infty)\right]$, where $h(r)$ is defined in eq. \ref{hr}.}
\label{rdep}
\end{figure}

  To understand the ``IR effect" of the finiteness of the box volume,  we performed calculations in $3$ different lattices, $64^3$, $80^3$ and $96^3$ , at fixed lattice spacing $v a=40/64$ ( as described above for the $64^3$ case). In Fig. \ref{rdep} we extrapolate these results to infinite volume using the function 
\be
h(r) &=& \int_ {-r+5} ^{r+5} dz\int _{-r}^rdy\int _{-r}^r dx\frac{1}{(x^2+y^2+z^2)^2},\label{hr}
\ee  
which is the integral of $1/r^4$ for a dyon sitting at $z=-5$, in a box of half width $r$. The infinity at the origin was removed since we only needed the long range behavior.
 The volume effect is found to be in agreement to the 
 expectation that the action density falls of as $1/r^4$.

\begin{figure}[h!]
\includegraphics[scale=0.49]{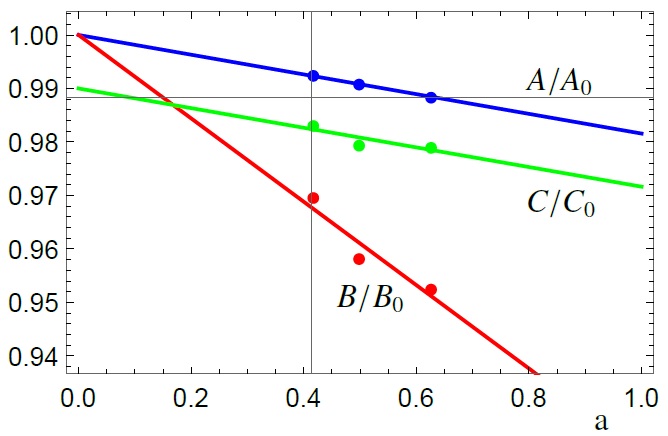}
\caption{The 3 parameters $A$, $B$ and $C$ normalized by their value at $a=0$ ($A_0$, $B_0$ and $C_0$) as a function of the lattice spacing $a$. We extrapolate the results to $a=0$ with a straight line. $C/C_0$ has been offset by $-0.01$ because it otherwise completely overlapped $A/A_0$. }
\label{adep}
\end{figure}

 In order to understand the ``UV effects" of discretization we also make 
 calculations for 3 different lattices, $64^3$, $80^3$ and $96^3$ ,  with variable lattice spacing but the same  volume (the same as  for $64^3$ point setting described  in section \ref{lattice_details}). In Fig. \ref{adep} we extrapolate these results to $a=0$ with a straight line.
 
\begin{figure}[h!]
\includegraphics[scale=0.48]{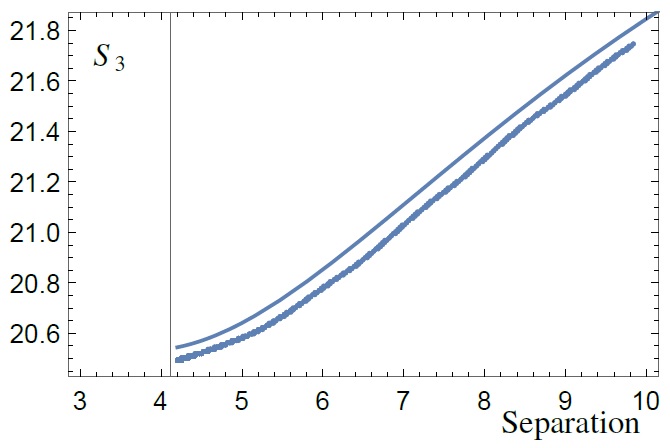}
\caption{
3d-action $S_3$ of the $M\bar{M}$ dyon pair vs separation between the dyons $|r_M-r_{\bar{M}} |v$ for $a=40/64$ for $96^3$ points (lower line) and for the extrapolated fit (upper line). The fit has been offset 
by $\Delta S_3= -0.9$ 
so its shape can be visually compared to the data. The separation of the dyons is defined by the maxima  of the action density. The  configuration  starts at the separation of $10/v$ between the dyons. The plot is terminated
after  the metastable configuration. The plot contains 1243 points (1 every 10 computer times) that makes up the fitted data. }
\label{shape}
\end{figure}

 Using those results we extrapolated to zero spacing and infinite box, by assuming that the two effects were independent.
 The resulting extrapolated curve is shown in Fig. \ref{shape} (upper curve, offset), to be compared to the actual data for the largest box (lower line). The offset value is basically the 5\% of the action outside the box mentioned earlier.
 This curve is  our main result.

We use the parameterization of the resulting curve of the type
\be S_3(r)=A\left(1-\frac{1}{r}+B\exp[-Cr]\right) , \label{eqn_param}\ee 
The reason the Coulomb part has no fitted parameter, other than the overall $A$,  is based on the following analytic argument.
 Classical interaction is known to be zero for interactions between two self-dual (or two  anti-self-dual) dyons due to the
 so called BPS protection. This means that the correction from the non-Abelian part  has to 
 cancel both electric and magnetic  Abelian Coulomb attractions.
 
  In the case we consider, the interaction of self-dual with anti-self-dual
 objects, the electric and magnetic Abelian Coulombs cancel each other, while the  non-Abelian part is expected to change its sign.
 
 We therefore expect the long range behavior for the action $S_3$ to behave like
 
\begin{eqnarray}
S_3(r \to \infty) &=& 8 \pi v + (m_1m_2-e_1 e_2)\frac{4\pi v}{r v},
\end{eqnarray} 
 
where $m_i$ and $e_i$ are the magnetic and electric charges of the two dyons and distance is given as $r v$ to show that all terms are proportional to $4\pi v$. The minus sign in front of the electric part, is due to the non-Abelian contribution. For the dyon-antidyon interaction we get  

\begin{eqnarray}
S_3(r \to \infty) &=& 8 \pi v(1-\frac{1}{rv}).
\end{eqnarray} 
 
 Since outcome of this argument was found to be in agreement with our numerical data, inside the errors, we decided not
 to include an extra parameter for the $1/r$ part. 
 The exponential was found to describe the potential nicely in the fitted region, while not affecting long range behavior.  
 We emphasize that our parameterization should only be used for $r\geq 4.2/v$.

The
values of the parameters and the errors obtained from the formal fit (to more than thousand points corresponding to
different separation during the gradient flow process) are
\be A=25.20\pm 0.01, \,   B=1.13\pm 0.03, \,
C=.607\pm .004 \ee
Note that he value of the parameter $A$ is only
 by 0.07 (or 0.3\%) higher than the action of two well separated dyons in continuum,  $S_3=8\pi \approx 25.13$.
 This fact confirms that our extrapolations are quite accurate, we did it as a test.
  In applications, one should of course use the analytic value of $A$ mentioned.
Note also that $B$ is about 30 standard deviations from zero, ensuring that an exponential term
is absolutely needed.

As a last comment in this section, we want to point out that the curve of the action $S_3$ never becomes completely flat at $r=4.2/v$ as otherwise expected. On the other hand, results with starting separations smaller than $r=4.2/v$ clearly converges to the same point around $r=4.2/v$. This indicates that the technique is not perfect around this point. This is most likely due to small changes to the shape of the action density, which were used to determine the position. These small changes could very well be caused by the collapse we always saw after enough time had passed. 

\subsection{Details of the Streamline}
In this section we focus on the properties  of the streamline configurations other than
the action. We will subsequently discuss: (i) How does the profile of the ``Higgs field" $A_4^3$ change. (ii) How does the charges change; and (iii)  What happens with the  Dirac strings. 

\subsubsection{The Higgs Field $A_4^3$}
Before we turn to the results, let us remind that for an individual dyon -- and thus for two at large separation --
the Higgs Field vanishes at the center. At large distances it should be the same value and direction: on the plots we
use a positive one. One might expect that the same shape will be maintained during the gradient flow on the streamline.

\begin{figure}[h!]
\includegraphics[scale=0.48]{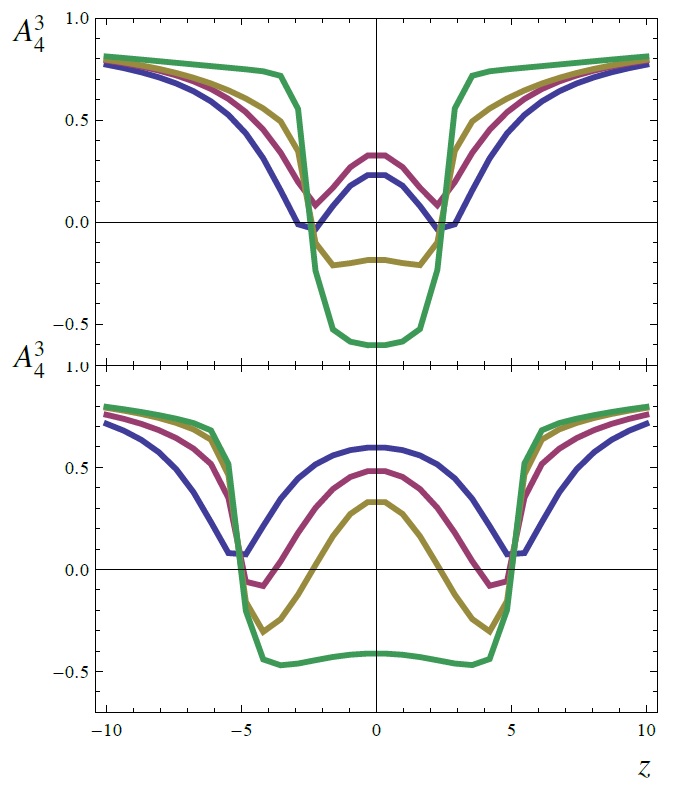}
\caption{Subsequent snapshots of $A^3_4$ along the z axis in natural units for an initial separation of $5/v$ (upper) and $10/v$ (lower) between the center of the 2 dyons. (upper) The configuration with the smallest field at the sides of the plotted area is the initial configuration. At computer time $\tau=5000$ the minimums have risen slightly, but is overall the same shape. At $\tau=9400$ the configuration has started collapsing. At $\tau=10000$ the configuration has collapsed to one minimum completely. (lower) The configuration with the smallest field at the sides of the plotted area is the initial configuration. At computer time $\tau=3000$ the minimums have moved slightly towards the middle and the minimums have become smaller. At $\tau=10000$ the configuration has reached the stable almost flat area in the action. At $\tau=14000$ the configuration has collapsed completely to an almost flat region between the initial positions of the dyons. }
\label{A34_comp_r5}
\end{figure}

As shown in Fig. \ref{A34_comp_r5}, this is not the case: the  Higgs Field goes through zero at the centers
and gets $negative$, about $-0.5v$, in between the dyons. The upper and lower plots are snapshots for 
 two different evolution histories, for an initial separation of $r=5/v, 10/v$, which show the same trend. 
 

\subsubsection{The Charges}
The electric and magnetic  charges inside certain sub-boxes are calculated via
the Gauss surface integrals. Total initial charges for the $M\bar{M}$ configuration should be $2$ for electric charges, and $0$ for magnetic charges. 

\begin{figure}[h!]
\includegraphics[scale=0.48]{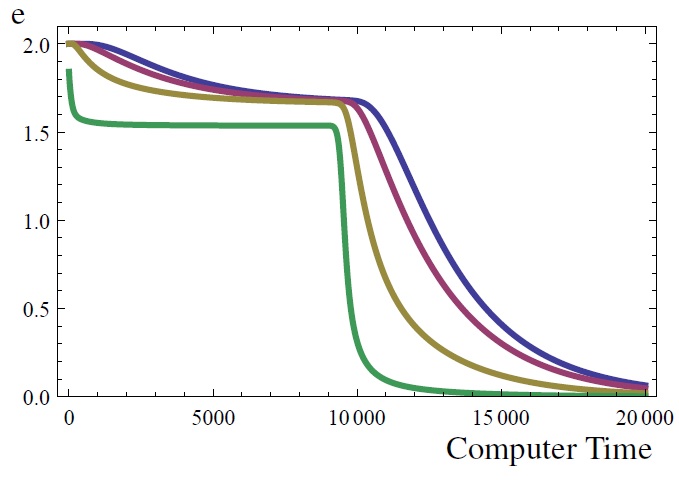}
\caption{Electric charge for $v = 1$ as a function of computer time for an initial separation $|r_M-r_{\bar{M}} |v=5$ between the dyons. The electric charge inside sub-boxes of width $38.75/v$, $28.75/v$, $18.75/v$ and $8.75/v$, centered at the origin. }
\label{vpi05rCharge}
\end{figure}

We further study charge locations using the  different sized boxes. These sub-boxes are all centered around the origin, with one dyon at $z=2.5/v$ and one dyon at $z=-2.5/v$. Total widths of the sub-boxes used are $38.75/v$, $28.75/v$, $18.75/v$ and $8.75/v$, while the width of the entire box used is $40/v$. Time evolution of the electric charge inside all sub-boxes is shown
in Fig. \ref{vpi05rCharge}. We observe that all charges are very stable for about 10000 time steps 
of the gradient flow, though we do see a small decrease in the total electric charge, after which 
the electric charge  quickly goes down  for all boxes. This happens at the same time as the action starts to drop as well. The fact that the smallest box shows zero sharply, while the
largest sub-box still contains about half of the charge 5000  time steps later, 
suggests that the electric charge moves out of the box gradually. 

\begin{figure}[h!]
\includegraphics[scale=0.48]{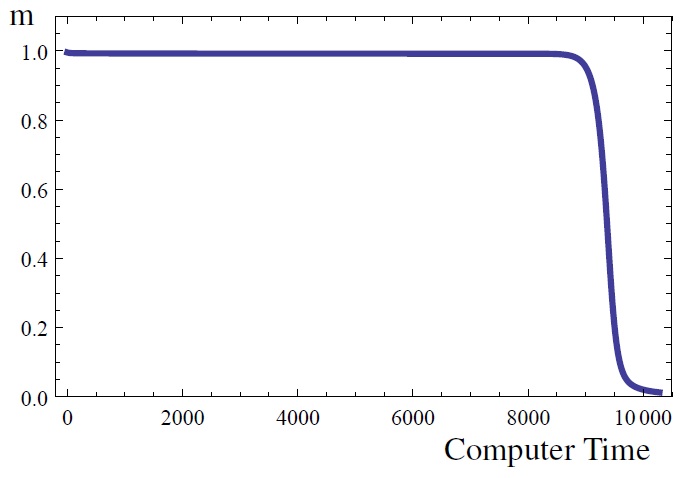}
\caption{Magnetic charge for $v = 1$ as a function of computer time for an initial separation $|r_M-r_{\bar{M}} |v=5$ between the dyons. The magnetic charge is found from a sub-box that goes from the middle of our lattice in $z$ (the dyons are separated along the z-axis) and to the edge, while filling up the entire part of the x and y-axis. The drop happens at the same time as the drop in action.}
\label{vpi05rMCharge}
\end{figure}

Since the total magnetic charge is zero, we 
cut each of the boxes described above in half in the xy-plane, through the origin. 
 That meant that only one dyon would be inside the sub-box.
 The time evolution of the magnetic charge inside the largest half-sub-box  is shown  in Fig. \ref{vpi05rMCharge}.
 The magnetic charge is very stable and close to 1, but collapses to $0$. The moment is the same as
 that for the action collapse.  We thus conclude that the magnetic structure is crucially important for the 
 preservation  of the individual solitons.

\subsubsection{The Dirac Strings}
\begin{figure}[h!]
\includegraphics[scale=0.5]{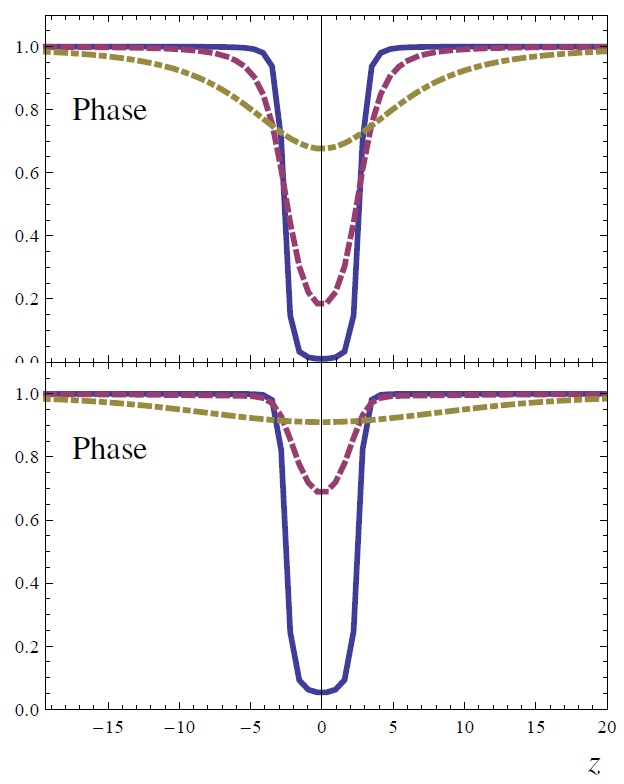}
\caption{The phase from the strings divided by $2\pi$ for a $M$ and $\bar{M}$ dyon at an initial separation $|r_M-r_{\bar{M}} |v=5$ along the z axis in natural units. (upper) Taken at the beginning of the simulation and (lower) Taken at the end of the simulation. The line is for a square loop with the sides of one link, the dashed is for 5 links and the dotdashed is for 21 links.}
\label{v1r5String}
\end{figure}

We now look at the Dirac strings. While those are  gauge transformation artifacts, we still wonder whether
the magnetic flux they carry is there or not, through the gradient flow process. 
 To observe the Dirac string we evaluate the phase of the spatial square loop $\oint dx_\mu A_\mu/(2\pi)$  winding around a  string as explained in section \ref{lattice}. 
 We plot the space of the spatial loop  along the z-axis for an initial configuration of $r=5/v$  in Fig. \ref{v1r5String} taken at the beginning (upper plot) and at the end of the process (lower plot),
 with loops of different size. 
 
For the smallest (square) loop  used  the phase takes a value close to 0 in between the two strings:  there is no string there. Increasing the size of the loop, the phase gets closer and close to $2\pi$, as expected. 
The pictures are very similar, and so the conclusion is that Dirac strings hardly change during the 
gradient flow process.

\subsection{$L\bar{L}$ pairs}
The $L\bar{L}$ pair has been studied in the gauge where the L dyons are constant in time. The overall result is the same as for the $M\bar{M}$ case and we therefore only point out the difference.

 While the $L$ and $\bar{L}$ dyons are time dependent in the gauge where $\langle A_4^3\rangle=v$, we can still explore the configuration in the time independent gauge before the time dependent gauge transformation is done. In the time independent gauge the Higgs field points in the negative direction with a value of $2\pi T -v$. To put $\langle A_4^3\rangle=v$ we need to do a time dependent gauge transformation, but this should not affect the results.

\begin{figure}[h!]
\includegraphics[scale=0.49]{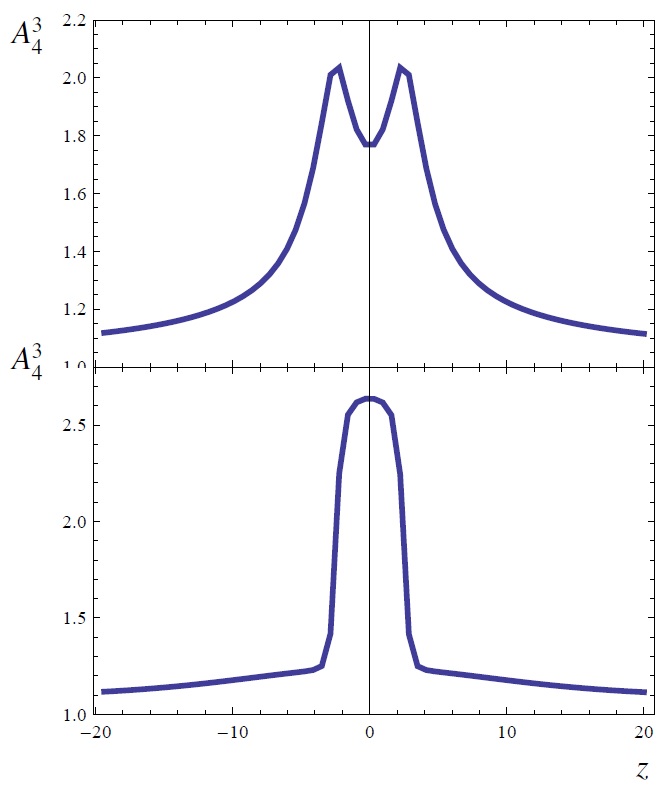}
\caption{$A_4 ^3$ for the $L\bar{L}$ dyon pair along the separation of the two dyons (z-axis), in natural units, at the beginning (upper) and end (lower) of the simulation for an initial separation of $|r_L-r_{\bar{L}} |v=5$. The dyons had $2\pi T -v =1$ and the time dependent gauge transformation has been performed to make the Higgs field at infinity equal to $1$.}
\label{vpi05r5HiggsStartLL}
\end{figure}

Since nothing different from the $M\bar{M}$ pairs happens for the charge and action we won't show those graphs. More interesting is the Higgs field which after a time dependent gauge transformation looks like in Fig. \ref{vpi05r5HiggsStartLL} for initial configuration (upper) and  for the configuration after the rapid drop in the action (lower).

\begin{figure}[h!]
\includegraphics[scale=0.48]{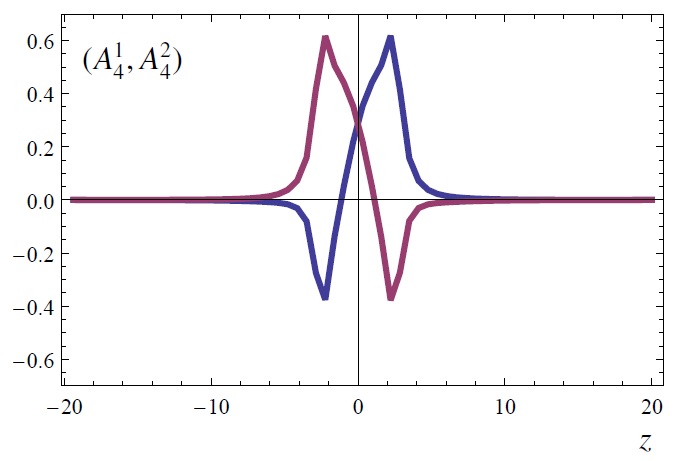}
\caption{$A_4 ^1$ and $A_4 ^2$ along the separation of the $L$ and $\bar{L}$ dyon (z-axis) for an initial separation of $|r_L-r_{\bar{L}} |v=5$, at the end of simulation, before the extra time dependent gauge transformation has been done.}
\label{vpi05r5A41A42EndLL}
\end{figure}

It is seen how the valleys are now instead a mountain for the $L$ and $\bar{L}$ dyons, since we have gauged the results such that $\langle A_4^3\rangle=1$.  After gauging back to the gauge where $\langle A_4^3\rangle=v$, we find that $A_4^3$ have gained a time dependent core. This time dependent core comes from the $\tau _1$ and $\tau _2$ component of $A _4$ which have become non-zero around the origin as shown in Fig. \ref{vpi05r5A41A42EndLL}. The $A_4^3$ component which is the mountain shown in Fig. \ref{vpi05r5HiggsStartLL} (lower), stays time independent.



When one do the time dependent gauge transformation, the time dependence that $A_4$ do gain is only for the $\tau _1$ and $\tau _2$ component of $A _4$, since $\exp(i\pi T x_4\tau_3)\tau _{(1,2)}=\tau _{(1,2)} \exp(-i\pi T x_4\tau_3)$. This means that the gauge transformation that puts $\langle A_4^3\rangle=v$ will mix the $\tau _1$ and $\tau _2$ component of $A _4$ with a time dependent phase.


\section{Summary and Outlook}
In summary, we performed the gradient flow studies of the instanton-dyon-antidyon configurations.
We found that, after a brief period of initial relaxation, the process settles in to a rather universal
``streamline" set of configurations, with steady and slow reduction of the action. We found that numerically
the dyon-antidyon interaction is relatively small. 

We found that  below a certain separation of the dyons the streamline no longer exists. 
We also found that between the  ``streamline" set and the final collapse to perturbative (small action) fields
there is  a very long-lived state at a separation of $4.2/v$, a dyon-antidyon molecule in which all forces are nearly exactly balanced. 
%
%
After certain time the magnetic charges annihilate each other and the configuration rapidly collapses to a pure gauge configuration, while the electric charge decreases in such a way, that it suggests that the charges propagate outside our box. For initial separations smaller than $4.2/v$, 
we find that configuration quickly moves into the meta-stable one mentioned, until they collapse. 

Our main conclusion is the universality of the streamline. We have performed zero spacing and infinitive volume
extrapolation of our results, and parameterize it to the form (\ref{eqn_param}), see parameters after the
formula. We believe the values of the parameters are quite accurate: those  constitute the main 
result of this work.


 


%

 Speaking about future work, one obvious extension of the current work would be a calculation of the Dirac eigenvalue spectrum 
for the streamline configurations at hand. Indeed, 
as experience with the ``instanton liquid"  has shown
\cite{Schafer:1996wv}, the main interaction in QCD-like theories is the fermion-induced
one, which happens precisely between the two duality sectors. 
It is even more so in the case of QCD-like theories with many quark flavors $N_f\sim 10$, which 
is currently under active investigation by the lattice community.
Recent work \cite{Shuryak:2012aa}
had argued that  this interaction leads to $L\bar{L}$ clusters of small size $\sim 1/N_f$,
a descendant of instanton-antiinstanton molecules. 
The reader may also find in this paper discussion of a number of specific issues/observables, relating 
the dyonic picture of QCD topology at $T>T_c$ to various lattice data.


The next logical step (after deriving the interaction between the topological objects in question) is
of course some study of the resulting statistical ensemble, improving on
simulations done in  \cite{Faccioli:2013ja}. Those studies are in progress and we hope to report the results shortly.

\vskip .25cm
 \textbf{Note added in proof: }
Since the paper was submitted, its results has been used in several studies of the
instanton-dyon ensembles. Liu, Shuryak and Zahed  \cite{Liu:2015ufa} argued that
dense enough ensemble is amenable to analytic mean field treatment, and had shown that
such ensemble is confining. They carried on to QCD-like theories with quarks  \cite{Liu:2015jsa}
and had shown that dense ensemble breaks chiral symmetry, provided the number of flavors and colors
satisfy $N_f < 2 N_c$. We pursued a direct numerical approach simulating numerically
ensembles of 64 and 128 dyons with variable densities, observing deconfinement phase transition 
\cite{Larsen:2015vaa} and chiral restoration in \cite{Larsen:2015tso} and concluding that
for $N_c=N_f=2$ QCD they occur at very similar densities.

We finally note that our main results can be  reformulated in modern terminology,
which quite recently came from mathematics to physics. 
Our finding of a near-stationary dyon-antidyon configuration at finite distance indicate existence of a
new extremal point of the path integral. The ``streamline" set of configurations
we found, 
going from it to a well separated pair, provides an example of a "Lefschetz thimble" gradient flow path 
connecting these two extrema.

\vskip .25cm \textbf{Acknowledgments.} \vskip .2cm 
This work is supported in part by the U.S. Department of Energy, Office of Science, under Contract No. DE-FG-88ER40388.

\end{document}